\documentclass[prl,superscriptaddress,twocolumn,showpacs,preprintnumbers,amsmath,amssymb]{revtex4}

\usepackage{graphicx}
\usepackage{dcolumn}
\usepackage{bm}
\newcommand{\czodcx}{$\rm Zn_xCu_{4-x}(OD)_6Cl_2$ }
\newcommand{\czodc}{$\rm ZnCu_{3}(OD)_6Cl_2$ }
\newcommand{\codc}{$\rm Cu_{2}(OD)_3Cl$}

\begin{document}

\preprint{}

\title{Magnetic field-induced instability of the cooperative paramagnetic state in Zn$_x$Co$_{4-x}$(OD)$_6$Cl$_2$}

\author{S. E. Dissanayake}
\affiliation{Department of Physics, University of Virginia, Charlottesville, VA 22904-4714, USA}
\author{C. Chan}
\affiliation{Department of Physics, Hong Kong University of Science and Technology, Clear Water Bay, Kowlon, Hong Kong}
\author{S. Ji}
\affiliation{Department of Physics, University of Virginia, Charlottesville, VA 22904-4714, USA}
\author{J. Lee}
\affiliation{Department of Physics, University of Virginia, Charlottesville, VA 22904-4714, USA}
\author{Y. Qiu}
\affiliation{NIST Center for Neutron Research, Gaithersburg, MD 20899, USA}
\affiliation{Department of Materials Science and Engineering, University of Maryland, College Park, MD 20742, USA}
\author{K. C. Rule}
\author{B. Lake}
\affiliation{Helmholtz Zentrum Berlin, GmbH, D-14109 Berlin, Germany}
\author{M. Green}
\affiliation{NIST Center for Neutron Research, Gaithersburg, MD 20899, USA}
\affiliation{Department of Materials Science and Engineering, University of Maryland, College Park, MD 20742, USA}
\author{M. Hagihala}
\affiliation{Department of Physics, Saga University, Saga 840-8502, Japan}
\author{X. G. Zheng}
\affiliation{Department of Physics, Saga University, Saga 840-8502, Japan}
\author{T. K. Ng}
\affiliation{Department of Physics, Hong Kong University of Science and Technology, Clear Water Bay, Kowlon, Hong Kong}
\author{S.-H. Lee}
\email{shlee@virginia.edu}
\affiliation{Department of Physics, University of Virginia, Charlottesville, VA 22904-4714, USA}

\date{\today}

\begin{abstract}
  Using elastic and inelastic neutron scattering techniques with and without application of an external magnetic field $H$, the magnetic ground states of Zn$_x$Co$_{4-x}$(OD)$_6$Cl$_2$ ($x=0,1$) were studied. 
Our results show that for $x=0$, the ground state is a magnetic long-range ordered (LRO) state where each tetrahedron forms an ``umbrella''-type structure.
On the other hand, for $x=1$, no static ordering was observed down to 1.5 K, which resembles the behavior found in the isostructural quantum system Zn$_x$Cu$_{4-x}$(OD)$_6$Cl$_2$.  When $H$ field is applied, however the $x=1$ system develops the same LRO state as $x=0$. This indicates that the $x=1$ disordered state is in the vicinity of the $x=0$ ordered state.
\end{abstract}

\maketitle


Recently, frustrated magnets where pairwise interactions cannot be satisfied simultaneously have been actively studied in the quest for novel quantum spin states in insulating materials \cite{balents10,pratt11,lee10,misguich05}. Among them, \czodcx~\cite{shores05} has generated particular interest because the magnetic Cu$^{2+}$ ions with quantum spin (s = 1/2)  form a two-dimensional kagome lattice  even though it has 10\% site-disorder. The interest was heightened when it was found that the system does not exhibit any magnetic ordering, short- or long-range, down to 50 mK for $x > 0.66$ \cite{helton07,mendels07,lee07}. The exact nature of the quantum disordered state is however controversial. It has been shown recently that in \czodc~there are broad gapless antiferromagnetic fluctuations that persist up to 20 meV \cite{vries09}. The wavevector- ($Q$)-dependence of the spin fluctuations was explained by a simple spin dimer model and interpreted as evidence that the ground state was an algebraic spin liquid state. On the other hand, when non-magnetic Zn atoms, which predominantly lie in triangular layers between the kagome planes, are replaced by the magnetic Cu atoms, long-range magnetic ordering sets in at low temperatures, for instance below $T_N = 6.7 $ K for \codc~ \cite{lee07}. In the LRO state, the magnetic excitation spectrum obtained from a powder sample showed a peak around the energy transfer of $\hbar\omega =$ 7 meV whose $Q$-dependence resembled that of spin dimers\cite{lee07} but may also be due to a van Hove singularity from the top of the spin-wave excitation band \cite{kim08}. It remains to be seen if the gapless spin fluctuations observed in \czodc~are characteristic of the quantum spin liquid state of the quantum kagome system \cite{vries08} or share the same origin as the $\hbar\omega =$ 7 meV mode of \codc. To address this issue, it is desirable to investigate other related materials that exhibit similar low temperature behaviors. A case in point is an isostructural compound, Zn$_x$Co$_{4-x}$(OD)$_6$Cl$_2$ with magnetic Co$^{2+}$ (3$d^7$; s = 3/2) ions. This system has a similar $x-T$ phase diagram as Zn$_x$Cu$_{4-x}$(OD)$_6$Cl$_2$. Co$_4$(OD)$_6$Cl$_2$ develops long-range magnetic order below $T_N = 10.5$ K which is suppressed upon doping with non-magnetic Zn while for ZnCo$_3$(OD)$_6$Cl$_2$, no static ordering has been observed down to 1.5 K.  

We have performed both elastic and inelastic neutron scattering measurements on powder samples of Zn$_x$Co$_{4-x}$(OD)$_6$Cl$_2$ with $x=0$ and $x=1$ in zero and non-zero external magnetic field, $H$. Our results are as follows. For $x = 0$ in the absence of an applied magnetic field the magnetic moments in the kagome plane order in a canted antiferromagnetic structure where their in-plane components form the $q=0$ 120$^{\circ}$ structure and these moments are canted out of the plane by 40$^{\circ}$. The magnetic moments in the triangular plane are aligned ferromagnetically along the $c$-axis. All moments have the same frozen moment of $\langle M \rangle = 3.77(3) \mu_{\textrm{B}}/{\textrm{Co}}^{2+}$ that is close to the expected value for the high-spin state of Co$^{2+}$ of 3.87 $\mu_{\textrm{B}}$. The ordered state exhibits three prominent dispersionless excitations centered at 3, 17, and 19 meV. The 17 meV mode is due to a magnetic crystal field splitting, while the 3 and 19 meV modes are spin waves. Our linear spin wave calculation shows that the canted AFM ground state and the two spin wave modes can be explained by antiferromagnetic nearest neighbor interactions in the kagome plane and relatively small ferromagnetic nearest neighbor interactions between the kagome and triangular planes. In addition a strong single ion anisotropy in the kagome plane is essential to reproduce the basic features of our data. For $x = 1$, without field ($H=0$) the system does not exhibit any magnetic order down to 1.5 K. Despite the lack of static order, it exhibits the 3 meV spin wave excitations found in the undoped compound.  When $H$ is applied, the system immediately develops the same LRO state as that of $x = 0$, and for $H > 3$ T the moment orders fully. Our results indicate that the disordered state of the $x=1$ system is in the vicinity of the ordered state of the $x = 0$ system, so that a weak external magnetic field can drive the disordered system into the ordered phase.

A 15 g powder sample of the $x = 0$ and a 1 g powder sample of the $x = 1$ system were grown using  hydrothermal solution reaction method described in Ref.  \cite{zheng06}. Neutron powder diffraction (NPD) measurements were performed on the BT1 powder diffractometer with Ge(311) monochromator ($\lambda$ = 2.0782 {\AA}), and the Rietveld refinement was carried out using the FULLPROF program. Inelastic neutron scattering measurements without an external magnetic field were carried out at the cold-neutron disk-chopper spectrometer (DCS) \cite{copley03} at the NIST Center for Neutron Research with $\lambda$ = 1.8 {\AA}, 2.5 {\AA} and 4.8 {\AA}. Elastic neutron scattering measurements under an external magnetic field were performed at the cold-neutron triple-axis spectrometer, V2, at the Helmholtz Zentrum Berlin with the energy of scattered neutrons fixed to $E_f$ = 5 meV. Contamination from higher-order neutrons was eliminated by using a Be filter. 

\begin{figure}
\includegraphics[width=0.48\textwidth]{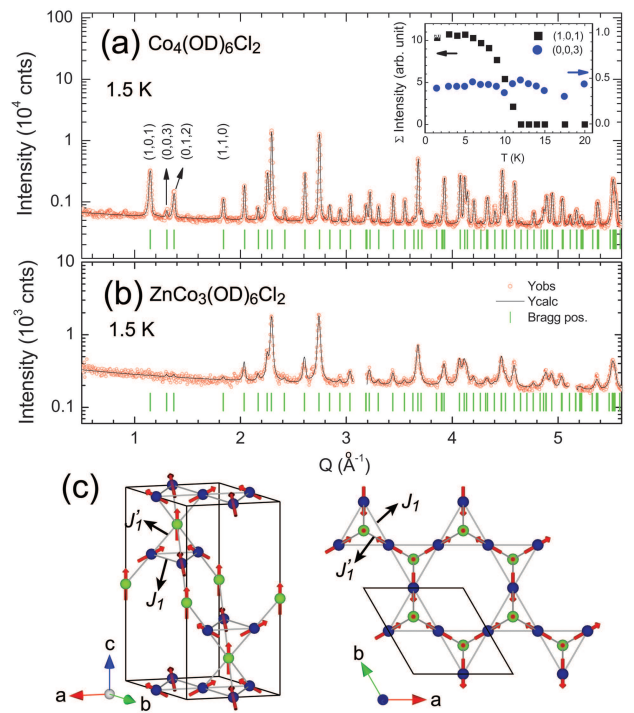}
\centering
\caption{(Color online) Neutron powder diffraction data of (a) Co$_4$(OD)$_6$Cl$_2$ 
and (b) ZnCo$_3$(OD)$_6$Cl$_2$ measured at 1.5 K below transition temperature, 
$T$ = 10.5 K. Circles are the data and the black line represents the calculated 
intensity obtained from the refinement by Fullprof. Green bars represent 
nuclear and magnetic Bragg peak position. Impurity peaks are excluded in (b).
The inset of (a) shows temperature dependence of integrated intensities of 
(1,0,1) and (0,0,3) peaks. (c) shows the schematic diagram of magnetic structure obtained from the 
refinement of Co$_4$(OD)$_6$Cl$_2$. Error bars in all figures represent 1 standard deviation from counting statistics.
\label{fig1}}
\end{figure}

\begin{table}
  \caption{The crystal structural parameters of Co$_4$(OD)$_6$Cl$_2$ and ZnCo$_3$(OD)$_6$Cl$_2$ obtained at 1.5 K by refining the data shown in Fig.\ref{fig1}~using the program FULLPROF. $B_{\textrm {iso}}$ is an isotropic thermal parameter expressed as $\exp(-B_{\textrm {iso}}\sin^2 \theta/\lambda^2)$, where $\theta$ is the scattering angle and $\lambda$ is the wavelength of the neutron.  \label{table1}}
  \begin{ruledtabular}
    \begin{tabular}{ccccc}
      \multicolumn{5}{l}{Co$_4$(OD)$_6$Cl$_2$ at 1.5 K, $R\bar3 m$,} \\
      \multicolumn{5}{l}{$\chi^2 = 1.35 $, R$_{wp}$ = 4.06, R$_{\textrm {Mag}}$ = 4.85} \\
\multicolumn{5}{l}{$a$ = $b$ = 6.8307(1) {\AA}, $c$ = 14.4473(3) {\AA}}\\ \hline
      Atom($W$) & $x$ & $y$ & $z$ & $B_{\textrm{iso}}$({\AA}$^2$) \\ 
      Co1 (3$b$) & 0 & 0 & 0.5 & 0.15(7) \\
      Co2 (9$e$) & 0.5 & 0 & 0 & 0.15(7) \\
      Cl (6$c$) & 0 & 0 & 0.2181(1) & 0.35(3) \\
      O (18$h$) & 0.2019(1) & -0.2019(1) & 0.0678(1) & 0.25(3) \\
      D (18$h$) & 0.2035(1) & -0.2035(1) & 0.5676(1) & 1.06(4) \\ \hline \hline
      \multicolumn{5}{l}{ZnCo$_3$(OD)$_6$Cl$_2$ at 1.5 K, $R\bar3 m$,} \\
      \multicolumn{5}{l}{$\chi^2 = 1.77 $, R$_{wp}$ = 6.82, } \\
\multicolumn{5}{l}{$a$ = $b$ = 6.8374(1) {\AA}, $c$ = 14.483(2){\AA}}\\ \hline
      Atom($W$) & $x$ & $y$ & $z$ & Occup. \\ 
      Co1 (3$b$) & 0 & 0 & 0.5 & 0.63(5) \\
      Co2 (9$e$) & 0.5 & 0 & 0 & 0.79(3) \\
      Zn1 (3$b$) & 0 & 0 & 0.5 & 0.37(5) \\
      Zn2 (9$e$) & 0.5 & 0 & 0 & 0.21(3) \\
      Cl (6$c$) & 0 & 0 & 0.2181(1) & 1 \\
      O (18$h$) & 0.2022(1) & -0.2022(1) & 0.0674(1) & 1 \\
      D (18$h$) & 0.2030(1) & -0.2030(1) & 0.5678(1) & 1 \\
    \end{tabular}
  \end{ruledtabular}
\end{table}

  Fig. \ref{fig1} shows neutron powder diffraction data of (a) Co$_4$(OD)$_6$Cl$_2$ and (b) ZnCo$_3$(OD)$_6$Cl$_2$ measured at 1.5 K.
  The lines are the refinement results of the nuclear and magnetic structures. For both systems, the crystal structure has an undistorted hexagonal structure with $R\bar{3}m$ symmetry, consistent with the previously reported structure for Co$_4$(OD)$_6$Cl$_2$. 
  The optimal parameters for their crystal structures are listed in Table I. Note that in 
  ZnCo$_3$(OD)$_6$Cl$_2$ the kagome lattice formed by the crystallographic $9e$ site is filled 79 \% by the Co$^{2+}$ ions, that is well above the percolation threshold for the lattice, $p_c =$ 0.62 \cite{scholl80}.
  Upon cooling, Co$_4$(OD)$_6$Cl$_2$ develops long-range magnetic order below $T_N = 10.5$ K (see the inset of Fig. 1 (a)), while ZnCo$_3$(OD)$_6$Cl$_2$ does not order down to 1.5 K. 
  The magnetic diffraction pattern of Co$_4$(OD)$_6$Cl$_2$ shown in Fig. 1 (a) is consistent with the previous work. On the other hand, our refinement showed that the magnetic structure is different from their ``2-1'' structure \cite{zheng06,zenmyo09}. 
  The ``2-1'' structure should yield magnetic contributions at (0 0 $L$) ($L = 3n$, $n$ is a integer), but as shown in the inset of Fig. 1 (a) those peaks do not show any increase below $T_N$. 
  The magnetic structure that yields the best refinement with the reliability factor of R$_{\textrm{Mag}}$ = 4.85 is an ``umbrella''-type antiferromagnetic structure as shown in Fig. 1 (c): 
  spins in the triangular plane (green balls) order ferromagnetically along the $c$-axis while kagome spins (blue balls) form a canted antiferromagnetic structure with $q=0$ where their $ab$-plane spin components have 120$^{\circ}$ arrangement with +1 chirality and their $c$-axis components are canted by 40$^{\circ}$ from the $ab$-plane \cite{inami00}.
  The frozen moment was determined to be $\langle M\rangle _{\textrm{Co}}$ = 3.77(3) $\mu_{\textrm{B}}/Co$ that is close to the expected value for the high-spin state of Co$^{2+}$ of 3.87 $\mu_{\textrm{B}}$.

\begin{figure}
\includegraphics[width=0.48\textwidth]{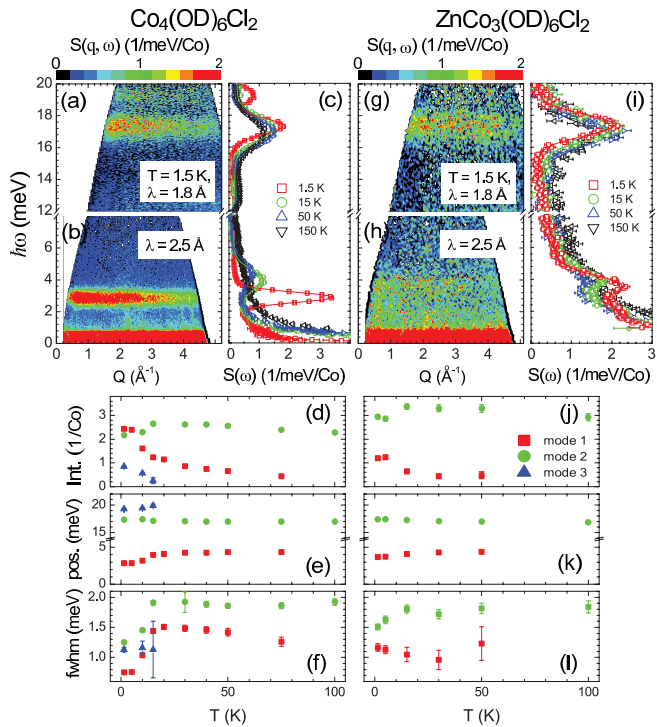}
\centering
\caption{(Color online) 
Inelastic neutron scattering data of Co$_4$(OD)$_6$Cl$_2$ ((a)-(f)) and ZnCo$_3$(OD)$_6$Cl$_2$ ((g)-(l)) measured at the DCS at NIST using three different wave lengths of incident neutron, $\lambda$ = 1.8 {\AA}, 2.5 {\AA} and 4.8 {\AA}.
(a) and (b): Countour maps of neutron scattering intensity $I(Q,\omega)$ as a function of energy transfer, $\hbar\omega$, and momentum transfer, $Q$, obtained at 1.5 K below $T_N$ = 10.5 K with $\lambda$ = 1.8 {\AA} and $\lambda$ = 2.5 {\AA}.
(c): neutron scattering intensities as a function of energy transfer measured at 1.5 K, 15 K, 50 K, and 150 K obtained from the integration of $I(Q,\omega)$ over 0.5 {\AA}$^{-1}$ $<$ $Q$  $<$ 4 {\AA}$^{-1}$.
 Solid lines are obtained from fitting by lorentzian functions.
(d)-(f): integrated intensity, peak position and fwhm of excitation modes as a function of temperature, respectively. 
(g)-(l) are plotted for ZnCo$_3$(OD)$_6$Cl$_2$ with the same manner of Co$_4$(OD)$_6$Cl$_2$.
\label{fig2}}
\end{figure}

  Let us now investigate the dynamic spin correlations in these systems. 
  Fig. 2 (a) and (b) show the contour map of neutron scattering intensity of Co$_4$(OD)$_6$Cl$_2$ as a function of energy transfer, $\hbar\omega$, and momentum transfer, $Q$, measured at $T$= 1.5 K.
  There are three prominent magnetic excitations centered at 2.9 meV, 17.3 meV and 19.3 meV. 
  All three excitations are resolution-limited in energy, and their intensities monotonically decrease as Q increases. This suggests that those modes are localized, probably due to frustration. 
  Another excitation centered at 14 meV becomes stronger as $Q$ increases, indicating it is not magnetic but due to lattice vibrations. We have performed similar measurements with $\lambda$ = 1.5 {\AA} and confirmed that there are no other magnetic excitations between 20 meV and 30 meV. 
  Upon warming, the three magnetic modes remain intact up to 11 K above which the 2.9 meV and 17.3 meV modes gradually shift to 4.0, 17.2 meV, respectively for $T > T_N$ (see Fig. 2 (c)).
  Upon further warming, the three modes slowly weaken but are present up to high temperatures: 75 K and 300 K for the 4.0 meV and 17.2 meV modes, respectively.
  This behavior is highly unusual for a magnetic system that undergoes long range order at low temperatures. Usually any spin wave excitation at low energies becomes quasi-elastic above $T_N$ while those at high energies remain up to a temperature that corresponds to their characteristic energy scale. Thus it is rather surprising that the 4.0 meV mode survives up to 75 K and the 17.2 meV all the way up to 300 K. 
  In comparison, Cu$_4$(OD)$_6$Cl$_2$ exhibits two prominent magnetic excitations centered at 1.3 meV and 7 meV in its LRO phase below 7 K.  Upon warming, the 1.3 meV mode shifts to lower energies, and becomes quasi-elastic above $T_N$ while the 7 meV mode remains at the same energy and gradually weakens then disappears above 18 K.

  Fig. 2(g)-(l) show the similar data taken from ZnCo$_3$(OD)$_6$Cl$_2$ that remains in a cooperative paramagnetic phase down to 1.5 K. Since the system does not order, one would expect quasi-elastic paramagnetic scattering, as found in many ordinary magnets as well as in its isostructural quantum magnet ZnCu$_3$(OD)$_6$Cl$_2$.
  On the contrary, at 1.5 K ZnCo$_3$(OD)$_6$Cl$_2$ exhibits two well-defined excitations centered at 3.7 and 17.4 meV in addition to quasi-elastic scattering that extends up to ~ 5 meV. The 3.7 and 17.4 meV modes are broader than those in Co$_4$(OD)$_6$Cl$_2$, and the 19.3 meV mode that was present in Co$_4$(OD)$_6$Cl$_2$ is absent.
  Upon warming, the 3.7 meV and 17.4 meV modes shift to 4.5 meV and 16.8 meV, respectively. 
  Upon further warming, the 4.5 meV mode weakens and disappears at 50 K, while the 16.8 meV mode weakens but exists up to 300 K.
  These behaviors closely resemble the T-dependence of the peak positions and intensities of the magnetic excitations in Co$_4$(OD)$_6$Cl$_2$, which indicates that the magnetic excitations in the two systems share the same origin even though their ground states are different. 
  It is noted that the integrated inelastic scattering intensities in dimension of $\mu_B^2$ at 1.5 K are 5.4(4) for Co$_4$(OD)$_6$Cl$_2$ and 12.1(8) for ZnCo$_3$(OD)$_6$Cl$_2$ comparable to the sum rule of $g^2S$ = 6 and $g^2S(S+1)$ = 15, respectively. 
  
\begin{figure}
\includegraphics[width=0.48\textwidth]{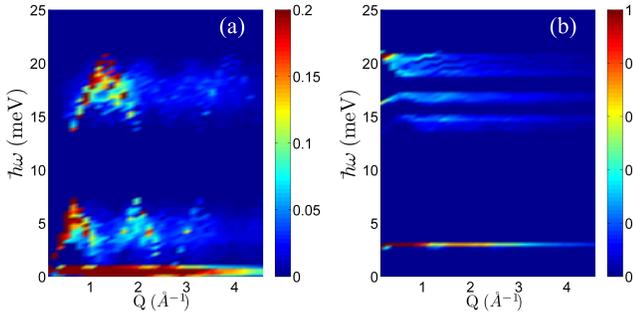}
\centering
\caption{(Color online) 
Linear spin wave calculation results - Contour map of powder-averaged neutron scattering cross section as a function of $Q$ and $\hbar\omega$. (a) $D_i<J_1$ (b) $D_i>>J_1$ for Co2 kagome spins. See the text for explanation.
\label{fig3a}}
\end{figure}

We performed linear spin wave calculations with the following effective minimal hamiltonian,
\begin{equation}
{\cal H} = \sum_{kag}J_1 {\bf S}_i \cdot {\bf S}_j + \sum_{kag-tri}J_1^{'} {\bf S}_i \cdot {\bf S}_j + \sum_{i}D_i ({\bf d}_i \cdot{\bf S}_i)^{2}  
\end{equation}
where $J_1$ is the nearest neighbor interaction in the kagome plane, $J_1^{'}$ represents the nearest neighbor interaction between the kagome and triangular plane and single ion anisotropy, $D_i$. The direction of the single ion anisotropy $d_i$, is along the $c$-axis for the triangular spins while it is perpendicular to the Co-Cl bond direction for the kagome spins.

We explore two cases of ${\cal H}$. Fig. 3(a) and 3(b) show the contour maps of powder-averaged neutron scattering cross section when the single ion anisotropy of the kagome Co2 spins is weak ($D_i<J_1$) and strong ($D_i>>J_1$), respectively.  For weak $D_i$ in the kagome spins, the 3 meV and 19 meV excitations can be reproduced by AFM interaction  $J_1$ = 1.15 meV, stronger FM interaction $J_1^{'}$ = -2.185 meV, single ion anisotropy of $D_i$ = -0.2 meV for the kagome Co2 spins and $D_i$ = -0.02 meV for the triangular Co1 spins, which stabilize in 45$^{\circ}$ canting angle. This model however shows two dispersionless modes centered at 0.3 meV and 0.6 meV, and dispersive modes centered around ~3 meV and ~19 meV, which are inconsistent with our data.

For strong $D_i$, a close match with the data can be obtained by AFM interaction  $J_1$ = 1.22 meV and relatively small FM interaction $J_1^{'}$ = -0.537 meV along with a strong single ion anisotropy of $D_i$ = -5.917 meV for the kagome Co2 spins and a weak single ion anisotropy $D_i$ = -0.05 meV for the triangular Co1 spins results in a canting angle of 40$^{\circ}$. As shown in Fig. \ref{fig3a}(b), the results explain quite well the two excitation modes centered at 3 meV and 19 meV. The mode at 17.2 meV in Fig. \ref{fig3a}(b) might indicate that the experimentally observed 17.2 meV mode is not entirely due to crystal field effect but has a contribution from spinwaves. The additional weak 15 meV mode was not observed experimentally.
 
\begin{figure}
\includegraphics[width=0.48\textwidth]{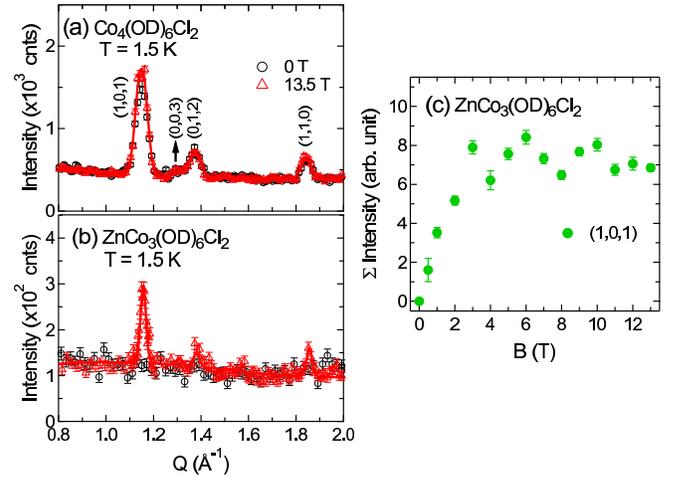}
\centering
\caption{(Color online) 
Elastic neutron scattering patterns of (a) Co$_4$(OD)$_6$Cl$_2$ and (b) ZnCo$_3$(OD)$_6$Cl$_2$ as a function of $Q$ up to 2.0 {\AA}$^{-1}$, obtained on V2 at $T$ = 1.5 K under external magnetic fields of 0 (black circle) and 13.5 T (red triangle). 
(c) Integrated intensity of (1,0,1) Bragg peak as a function of external magnetic field from 0 to 13.5 T for ZnCo$_3$(OD)$_6$Cl$_2$.
\label{fig4}}
\end{figure}

  In order to shed light on the nature of the ground states and the observed magnetic excitations of both compounds, we have performed elastic neutron scattering measurements under an external magnetic field upto 13.5 T. 
  Fig. 4 (a) and (b) show elastic neutron scattering data obtained from Co$_4$(OD)$_6$Cl$_2$ and ZnCo$_3$(OD)$_6$Cl$_2$, respectively, as a function of $Q$, measured at $T$ = 1.5 K with $H$ = 0 T and $H$ = 13.5 T.
 For Co$_4$(OD)$_6$Cl$_2$, the field does not induce any change in the elastic scattering pattern.
 On the other hand, for ZnCo$_3$(OD)$_6$Cl$_2$ the field higher than 3 T induces the same magnetic long range ordering as that of Co$_4$(OD)$_6$Cl$_2$. (see Fig. 4 (b) and (c)) 
 This indicates that the cooperative paramagnetic ground state of ZnCo$_3$(OD)$_6$Cl$_2$ is very close to the LRO state.

 The similar dynamic spin correlations in Zn$_x$Co$_{4-x}$(OD)$_6$Cl$_2$ both with $x=0$ and $x=1$, and the magnetic field induced instability of the paramagnetic ground state of $x=1$ tell us that a similar physics may occur in its quantum sister material; the quantum disordered state of ZnCu$_3$(OD)$_6$Cl$_2$ might be in a close proximity to the ordered state of Cu$_2$(OD)$_3$Cl. 
 Indeed, a very recent NMR study on ZnCu$_3$(OH)$_6$Cl$_2$ revealed that an external magnetic field could drive the quantum spin liquid (QSL) state into a spin-solid \cite{jeong11}. Our results on the Co-based material suggest that the $H$-induced spin-solid might be similar to the long-range ordered state of Cu$_2$(OD)$_3$Cl. This proximity scenario may also explain the recent neutron scattering results of ZnCu$_3$(OD)$_6$Cl$_2$
that showed the magnetic excitation continuum of the QSL state had a similar Q-dependence \cite{vries09} as that of the high energy spin-wave mode found in the ordered state of Cu$_2$(OD)$_3$Cl \cite{lee07,kim08}.

We thank M.-H. Whangbo and H.-J. Koo for helpful discussions. Research at UVA was supported by the U.S. Department of Energy, Office of Basic Energy Sciences, Division of Materials Sciences and Engineering under Award Nos. DE-FG02-07ER46384. TKN and CC acknowledge support from HKRGC through GRF grant 603410. The work at NIST was supported in part by the National Science Foundation under Agreement No. DMR-0944772.

\end{document}